\documentclass[prl,twocolumn,showpacs]{revtex4}
\usepackage[dvips]{graphicx}
\usepackage{latexsym}
\usepackage{amsmath}
\usepackage{amsfonts}
\usepackage{amssymb}
\usepackage{bm}
\usepackage{color}
\begin{document}
\newcommand{\fig}[2]{\includegraphics[width=#1]{#2}}
\newcommand{{\vhf}}{\chi^\text{v}_f}
\newcommand{{\vhd}}{\chi^\text{v}_d}
\newcommand{{\vpd}}{\Delta^\text{v}_d}
\newcommand{{\ved}}{\epsilon^\text{v}_d}
\newcommand{{\vved}}{\varepsilon^\text{v}_d}
\newcommand{{\up}}{\uparrow}
\newcommand{{\down}}{\downarrow}
\newcommand{{\bk}}{{\bf k}}
\newcommand{{\bq}}{{\bf q}}
\newcommand{{\tr}}{{\rm tr}}
\newcommand{\pprl}{Phys. Rev. Lett. \ }
\newcommand{\pprb}{Phys. Rev. {B}}

\title{Asymptotic solution of the Hubbard model in the limit of large coordination: Doublon-holon binding, Mott transition, and fractionalized spin liquid}
%
\author{Sen Zhou$^1$, Long Liang$^2$, and Ziqiang Wang$^{3}$}
\affiliation{$^1$ CAS Key Laboratory of Theoretical Physics,
Institute of Theoretical Physics, Chinese Academy of Sciences, Beijing 100190, China}
\affiliation{$^2$ COMP Centre of Excellence, Department of Applied Physics, Aalto University School of Science, FI-00076 Aalto, Finland}
\affiliation{$^3$ Department of Physics, Boston College, Chestnut Hill, MA 02467, USA}

\date{\today}
\begin{abstract}

An analytical solution of the Mott transition
is obtained for the Hubbard model on the Bethe lattice in the large coordination number ($z$) limit. The excitonic binding of doublons (doubly occupied sites) and holons (empty sites) is shown to be the origin of a continuous Mott transition between a metal and an emergent quantum spin liquid insulator.
The doublon-holon binding theory enables a different large-$z$ limit and a different phase structure than the dynamical meanfield theory by allowing intersite spinon correlations to lift the $2^N$-fold degeneracy of the local moments.
We show that the spinons are coupled to doublons/holons by a dissipative compact U(1) gauge field that is in the deconfined phase, stabilizing the spin-charge separated gapless spin liquid Mott insulator.

\typeout{polish abstract}
\end{abstract}


\maketitle

A Mott insulator is a fundamental quantum electronic state protected by a nonzero energy gap for charge excitations that is driven by Coulomb repulsion but not associated with symmetry breaking \cite{Mottbook}. It differs from the other class of insulators and magnets, better termed as Landau insulators, that require symmetry breaking order parameters produced by the residual quasiparticle (QP) interactions in a parent Fermi liquid. The most striking feature of Mott insulators is the separation of charge and spin degrees of freedom of the electron that completely destroys coherent QP excitations. A ubiquitous example of a Mott insulator is the quantum spin liquid where the spins are correlated but do not exhibit symmetry-breaking long-range order \cite{anderson73,wen91,palee08,balents10}. The spin liquid states have been observed in the $\kappa$-organics {\em near} the Mott metal-insulator transition \cite{kanoda03,kanoda05,kanoda08,matsuda08}.
The Mott insulator and the Mott transition are at the heart of the strong correlation physics. It is conceivable that the Mott insulator is the ultimate parent state of strong correlation from which many novel quantum states emerge 
\cite{anderson87,kivelson,phillips,leermp,weng}.

In this work, we provide a theory for the Mott insulator and the Mott transition in the Hubbard model at half-filling. The Hilbert space here is local and consists of doubly occupied (doublon), empty (holon), and singly occupied (spinon) states. The electron spectral function in different scenarios of the Mott transition is sketched in Fig.~1.
Focusing exclusively on the coherent QP, Gutzwiller variational wave function approaches \cite{gutzwiller} obtained a strongly correlated Fermi liquid \cite{dv-rmp84} that undergoes
a Brinkman-Rice (BR) transition \cite{br} to a ``pathological'' localized state with vanishing QP bandwidth and vanishing doublon (D) and holon (H) density (Fig.~1a). The dynamical meanfield theory (DMFT) maps the Hubbard model to a quantum impurity embedded in a self-consistent bath \cite{dmftrmp96,dv2012}. The mapping is exact only in a well-defined large-$z$ limit. The obtained $T=0$ Mott transition shown in Fig.~1b shows that the opening of the Mott gap at $U_{c1}$ and the disappearance of the QP coherence at $U_{c2}$ do not coincide such that the QP states in the metallic state for $U_{c1}<U<U_{c2}$ are separated from the incoherent spectrum by a preformed gap. This peculiar property \cite{nozieres,kehrein,gebhard} was shown to be correct \cite{kotliar} for the large-$z$ limit taken in DMFT where the spin-exchange interaction $J\sim t^2/U$ scales with $1/z$ and forces the insulating state to be a local moment phase with $2^N$-fold degeneracy. The present theory builds on a different asymptotic solution on the Bethe lattice using the slave-boson formulation to capture the most essential Mott physics, i.e. the excitonic binding between oppositely charged doublons and holons \cite{kaplan,yokoyama,capello,leigh,zhouwangwang,mckenzie}. We construct a different large-$z$ limit than DMFT and find a continuous Mott transition shown in Fig.~1c from a correlated metal to an insulating quantum spin liquid, where the opening of the Mott gap and the vanishing of the QP coherence coincide at the same $U_c$. The BR transition is preempted by quantum fluctuations and replaced by the Mott transition.
A key feature of our asymptotic solution is that on the insulating side of the Mott transition, while all doublons and holons are bound in real space into excitonic pairs with the Mott charge gap set by the D-H binding energy \cite{zhouwangwang}, the spinon intersite correlations remain and survive the large-$z$ limit, forming a gapless spin liquid by lifting the ground state degeneracy. We derive the compact gauge field action in the large-$z$ limit and show that the emergent dissipative dynamics drives the gauge field to the deconfinement phase where the fractionalized U(1) spin liquid is stable.
\begin{figure}
      \begin{center}
    \fig{3.2in}{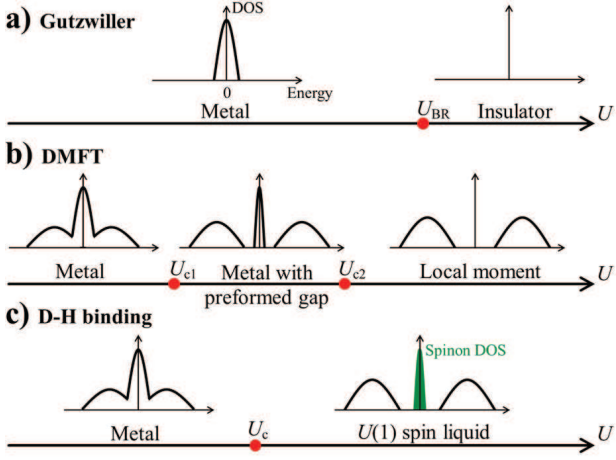}\caption{Schematic diagram of the Mott transition in (a) Gutzwiller with $U_{\rm BR}\simeq3.40D$ (a); (b) DMFT with $U_{c1}\simeq2.38D$ and $U_{c2}\simeq3.04D$ \cite{dmftrmp96,karski,dv2012}; and (c) present D-H binding theory with $U_c\simeq0.8U_{\rm BR}=2.71D$. $D$ is half bandwidth.
    \label{fig0}}
    \end{center}
    \vskip-0.5cm
    \end{figure}

Consider the Hubbard model on the Bethe lattice
\begin{equation}
H = -t \sum_{\langle ij \rangle}c_{i\sigma }^ \dagger  c_{j\sigma }  + {\rm h.c.} + U\sum_i {n_{i \uparrow } n_{i \downarrow } },
\label{h}
\end{equation}
where the $t$-term describes electron hopping on $z$ nearest neighbor bonds and the $U$-term is the on-site Coulomb repulsion. To construct a strong-coupling theory that is nonperturbative in $U$, Kotliar and Ruckenstein \cite{kr} introduced a spin-1/2 fermion $f_\sigma$ and four slave bosons $e$ (holon), $d$ (doublon), and $p_{\sigma}$ to represent the local Hilbert space for the empty, doubly-occupied, and singly occupied sites respectively: $\vert0\rangle=e^\dagger \vert\text{vac}\rangle$,  $|\!\!\uparrow\downarrow\rangle=d^\dagger f_\downarrow^\dagger f_\uparrow^\dagger \vert\text{vac}\rangle$, and $\vert \sigma\rangle= p_\sigma^\dagger f_{\sigma}^\dagger \vert\text{vac}\rangle$. The physical Hilbert space obtains under the holomorphic constraints for completeness
$e_i^\dagger  e_i + \sum_\sigma p_{i\sigma}^\dagger  p_{i\sigma}  + d_i^\dagger d_i  =1$ and the consistency of particle density $f_{i\sigma}^\dagger f_{i\sigma}=p_{i\sigma}^\dagger p_{i\sigma}+d_i^\dagger d_i$. The Hubbard model is thus faithfully represented by
\begin{equation}
H=-t\sum_{\langle ij \rangle} Z_{i\sigma}^\dagger Z_{j\sigma} f_{i\sigma}^\dagger f_{j\sigma} + {\rm h.c.} +
U\sum_i d_i^\dagger d_i,
\label{hslaveboson}
\end{equation}
where
$Z_{i\sigma}=L_{i\sigma}^{-1/2}(p_{i\bar\sigma}^\dagger d_i+e_i^\dagger p_{i\sigma})R_{i\bar\sigma}^{-1/2}.
$
The operators $L_{i\sigma}=1-d_i^\dagger d_i-p_{i\sigma}^\dagger p_{i\sigma}$ and $R_{i\bar\sigma}=1-e_i^\dagger e_i-p_{i\bar\sigma}^\dagger p_{i\bar\sigma}$ should be understood as projection operators
and the choice of the $-1/2$ power reproduces the results of the Gutzwiller approximation at the meanfield level \cite{kr}; although it can be argued to arise from the hardcore nature of the slave bosons. To study the Mott transition and the Mott insulating state, we focus on the spin SU(2) symmetric phases of the Hubbard model.

When the quantum states in Eq.~(1) are spatially extended,
the nearest neighbor single-particle correlator scales as $\langle c_{i\sigma}^\dagger c_{j\sigma}\rangle\sim 1/\sqrt{z}$. Hence the hopping $t$ must be rescaled according to $t\to t/\sqrt{z}$ in order to maintain a finite kinetic energy
in the large-$z$ limit \cite{metzner89}, such as in DMFT \cite{gk92}. While natural for the metallic phase, on the Mott insulating side the exchange interaction $J \sim t^2/U$ is forced to scale according to $J \to J/z$ which suppresses the intersite correlations that may otherwise lift the $2^N$-fold degeneracy in the ground state \cite{dmftrmp96}. This ultimately leads to an immediate emergence of the local moments on the insulating side and eliminates the possibility of a quantum spin liquid
where charges are localized but spins form a correlated liquid state.

The slave-boson formulation of the Hubbard model in Eq.~(\ref{hslaveboson}) offers a different large-$z$ limit 
without invoking the rescaling of $t$. This can be seen intuitively since the hopping of spinons in Eq.~(\ref{hslaveboson}) between neighboring sites is always accompanied by the co-hopping of the $Z$-bosons and vice versa. Thus, the effective hopping amplitude of the spinon/$Z$-boson carries a dynamically generated $1/\sqrt{z}$ from the $Z$-boson/spinon intersite correlator, resulting in a finite total kinetic energy in the large-$z$ limit. A spin liquid can thus emerge on the Mott insulating side. We note that on the metallic side not discussed in detail here, the D/H condensate contributions need to be scaled by $1/\sqrt{z}$ in the intersite correlator in order to be treated on equal footing as the uncondensed part to keep the kinetic energy finite, in close analogy to the recent formulation of the bosonic DMFT \cite{bdmft}. We focus on the Mott insulator at $U>U_c$, where all holons ($e$) and doublons ($d$) are bound into exciton pairs. 
Although the doublon and holon densities are nonzero ($n_d=n_e\neq0$), their single-particle condensates are absent.
To leading order in $1/z$, the single-occupation $p_{\sigma}$ bosons must therefore condense, i.e. $p_{\sigma}=p_\sigma^\dagger=p_0$. The operators $L_{i\sigma}$ and $R_{i\sigma}$ that enter $Z_{i\sigma}$ cannot introduce additional intersite correlations and must depend only on the local densities. This leads to $L_{i\sigma}=R_{i\sigma}=1/2$ at half-filling and $Z_{i\sigma}=2p_0(d_i+e_i^\dagger)$. Thus
$\langle Z_{i\sigma}^\dagger Z_{j\sigma}\rangle \sim 1/\sqrt {z}$. Together with $\langle f_{i\sigma}^\dagger f_{j\sigma}\rangle\sim 1/\sqrt{z}$, the electron correlator $\langle c_{i\sigma}^\dagger c_{j\sigma}\rangle\sim 1/z$ and the kinetic energy is finite in the large-$z$ limit.

The Hamiltonian in Eq.~(\ref{hslaveboson}) becomes,
\begin{eqnarray}
H=&-&4p_0^2t\sum_{\langle ij \rangle}\bigl[(d_i^\dagger d_j+e_j^\dagger e_i+e_i d_j +d_i^\dagger e_j^\dagger ) f_{i\sigma}^\dagger f_{j\sigma}
\nonumber \\
&+&{\rm h.c.}\bigr]
+ U\sum_i d_i^\dagger d_i.
\label{h-largez}
\end{eqnarray}
The condensation of the $p_\sigma$ bosons collapses two of the constraints to $n_d+p_0^2=n_\sigma^f$. The remaining one can be written as $e_i^\dagger e_i-d_i^\dagger d_i+\sum_\sigma f_{i\sigma}^\dagger f_{i\sigma}=1$, which corresponds to the unbroken $U(1)$ gauge symmetry and
specifies the gauge charges of the particles. Thus, increasing the spinon number by one must be accompanied by either destroying a holon or creating a doublon at the same site in the Mott insulator. The partition function can be written down as an imaginary-time path integral
\begin{equation}
Z=\int {\cal D}[f^\dagger,f]{\cal D}[d^\dagger,d]{\cal D}[e^\dagger e]{\cal D}[a_0,a] {\cal D}\lambda e^{-\int_0^\beta {\cal L} d\tau}. 
\label{partition}
\end{equation}
The Lagrangian is given by
\begin{eqnarray}
{\cal L}&=&\sum_i [f_{i\sigma}^\dagger(\partial_\tau+ia_0)f_{i\sigma}+
d_i^\dagger(\partial_\tau -ia_0)d_i+e_i^\dagger(\partial_\tau
\nonumber \\
&+&ia_0)e_i]+i\sum_i\lambda_i(d_i^\dagger d_i+ e_i^\dagger e_i+2p_0^2-1)-H,
\label{lagrangian}
\end{eqnarray}
where the Hamiltonian $H=H_f+H_b$,
\begin{eqnarray}
H_f=&-&{t_f\over\sqrt{z}}\sum_{\langle i,j\rangle}(e^{ia_{ij}}f_{i\sigma}^\dagger f_{j\sigma}+{\rm h.c.})
\label{hf} \\
H_b=&-&{t_b\over\sqrt{z}}\sum_{\langle i,j\rangle}\bigl[e^{-ia_{ij}}(e_j^\dagger e_i+d_i^\dagger d_j
\label{hb} \\
&+&e_id_j+d_i^\dagger e_j^\dagger)+{\rm h.c.}\bigr]+{U\over2}\sum_i(d_i^\dagger d_i+e_i^\dagger e_i),
\nonumber
\end{eqnarray}
with $t_f=8tp_0^2\sqrt{z}(\chi_d+\Delta_d)$, $t_b=8tp_0^2\sqrt{z}\chi_f$. In a stationary state, $\chi_d=\langle d_i^\dagger d_j\rangle=\langle e_j^\dagger e_i\rangle$ is the quantum average of the D/H nearest neighbor hopping, $\chi_f=\langle f_{i\sigma}^\dagger f_{i\sigma}\rangle$ the fermion hopping per spin, and $\Delta_d=\langle d_i^\dagger e_j^\dagger\rangle=\langle e_id_j\rangle$ is the D-H binding order parameter. In Eqs.~(\ref{lagrangian}-\ref{hb}), the spinons and the D/H are coupled by the emergent U(1) gauge fields $a_0$ and $a_{ij}$ associated with the constraint.
Physically, the instantons of this compact gauge field correspond to the tunneling events where the spinons and D/H tunnel in and out of the lattice sites \cite{ioffelarkin}.

We will first obtain the stationary state solution with $a_0=a_{ij}=0$, and then study the properties of the gauge field fluctuations. Eq.~(\ref{hf}) shows that the spinon hopping amplitude is $t_f/\sqrt{z}$ where $t_f$ scales with the D/H density, resulting in a bandwidth on the order of exchange coupling $J\sim t^2/U$. The spinon kinetic energy per site is $K_f=(t_f/ t)K_0$ where $K_0=2\int_0^D\rho_0(\omega)\omega d\omega$ is that for noninteracting electrons with hopping $t/\sqrt{z}$ and $\rho_0$ is the corresponding semicircle density of states $\rho_0(\omega)={2\over \pi D}\sqrt{1-(\omega/D)^2}$ on the infinite-$z$ Bethe lattice with a half-bandwidth $D=2t$. Thus, $K_0={4D/ 3\pi}$ and $K_f=8t_f/3\pi$.
Alternatively, expressing $K_f=4t_f\sqrt{z}\chi_f$, we obtain $\chi_f={1\over\sqrt{z}}{2\over3\pi}$. The effective boson hopping in Eq.~(\ref{hb}) is thus $t_b=8p_0^2t\sqrt{z}\chi_f=16p_0^2t/3\pi$ which is of the order $t$. The spectrum of the charge excitations residing in the D/H sector has a bandwidth on the order of the bare electron bandwidth, representing the large incoherent spectral weight in the Mott insulator.

From Eqs (\ref{lagrangian}) and (\ref{hb}), the stationary state bosonic Hamiltonian in the D/H sector is
\begin{equation}
{H}_{\rm D/H} = \int_{-D}^D d\omega\rho_0(\omega)\left( d_\omega^\dagger,e_\omega \right) \left(
\begin{array}{cc} \varepsilon_\omega & - \Delta_\omega \\
-\Delta_\omega & \varepsilon_\omega \end{array} \right)
\left( \begin{array}{c} d_\omega \\ e_\omega^\dagger
\end{array} \right),
\nonumber
\end{equation}
where $\varepsilon_\omega={U\over2}+\lambda-{t_b\over t}\omega$, $\Delta_\omega={t_b\over t}\omega$ are the D/H kinetic and pairing energies; $\lambda=\langle i\lambda\rangle$. Diagonalizing $H_{\rm D/H}$ using the Bogoliubov transformation produces two degenerate branches for the D/H excitations: $\Omega_\omega=\sqrt{\varepsilon_\omega^2-\Delta_\omega^2}$. The Mott insulator is thus an excitonic insulator and the Mott gap is given by the charge gap in $\Omega_\omega$,
\begin{equation}
G_{\rm Mott}(U)=2\Omega_D=2\sqrt{\left({U\over2}+\lambda\right)
\left({U\over2}+\lambda-4t_b\right)}.
\label{mottgap}
\end{equation}
The physical condition for a real $\Omega$ requires 
$U\ge 8t_b-2\lambda$ and the equal sign determines the critical $U_c$ for the Mott transition where $G_{\rm Mott}(U_c)=0$.
\begin{figure}
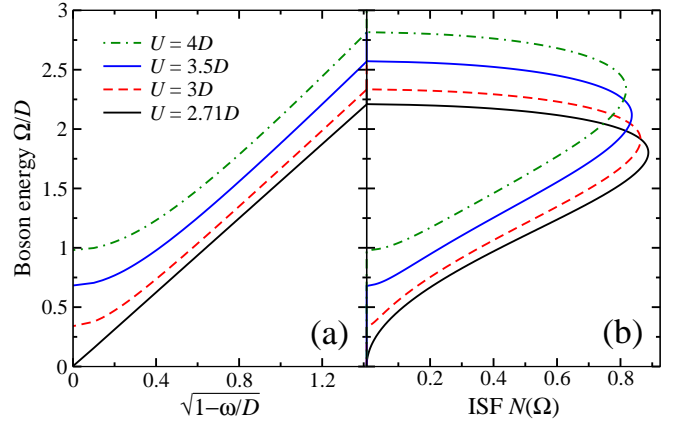

      \begin{center}
    \fig{3.4in}{fig2.eps}\caption{The doublon/holon energy spectrum (a) and the corresponding spectral density of state (b) for different $U$. \label{fig1}}
    \end{center}
    \vskip-0.65cm
    \end{figure}

Minimizing the energy leads to the self-consistent equations, $p_0^2={1\over2}-n_d$, $\lambda=4K_0\sqrt{z}(\chi_d+\Delta_d)$, and
\begin{eqnarray}
n_d&=&{1\over2}\int_{-D}^{D}\left({\varepsilon_\omega\over\Omega_\omega}
-1\right)\rho_0(\omega)d\omega,
\label{nd} \\
\chi_d&=&{1\over2D\sqrt{z}}\int_{-D}^D {\varepsilon_\omega\over\Omega_\omega}\omega\rho_0(\omega)d\omega,
\label{chid} \\
\Delta_d&=&{1\over2D\sqrt{z}}\int_{-D}^D {\Delta_\omega\over\Omega_\omega}\omega\rho_0(\omega)d\omega.
\label{deltad}
\end{eqnarray}
Eq.~(\ref{nd}) shows that the nonzero D/H density is entirely due to quantum fluctuations above the Mott gap in $\Omega_\omega$ for $U>U_c$. Lowering $U$ toward $U_c$, $G_{\rm Mott}$ must reduce to host the increased D/H density until $G_{\rm Mott}=0$ at $U=U_c$ where the D/H condensation emerges and the continuous Mott transition takes place (Fig.~1c).

The D/H excitation spectrum is plotted in Fig.~2(a), showing the closing of the Mott gap as $U$ is reduced toward $U_c$. Note that the spectral density in Fig.~2(b) vanishes quadratically upon gap closing, which ensures that the Mott transition is continuous at zero temperature. Remarkably, the critical properties of the transition can be determined analytically. First, using the expression for $\lambda$, the critical $U_c$ is obtained from Eq.~(\ref{mottgap}),
\begin{equation}
U_c=U_{\rm BR}[1-2n_d^c-\sqrt{z}(\chi_d^c+\Delta_d^c)],
\label{uc}
\end{equation}
where $U_{\rm BR}=8K_0=32D/3\pi$ is the critical value for the BR transition on the Bethe lattice.
Eq.~(\ref{uc}) shows that the Mott transition emerges as the quantum correction to the BR transition due to D-H binding. Since $U_c<U_{\rm BR}$, the BR transition is pre-emptied by the Mott transition. At $U=U_c$, the D/H excitation spectrum becomes $\Omega_\omega=8K_0p_0^2\sqrt{1-\omega/D}$ as in Fig.~2(a), which is independent of $\chi_d$ and $\Delta_d$. The integrals in Eqs.~(\ref{nd}-\ref{chid}) can all be evaluated analytically to obtain the critical doublon density $n_d^c=(12\sqrt{2}-5\pi)/10\pi\simeq0.040$, D/H hopping $\sqrt{z}\chi_d^c=2\sqrt{2}/35\pi\simeq0.026$, and D-H binding $\sqrt{z}\Delta_d^c=22\sqrt{2}/105\pi\simeq0.094$. The critical value for the Mott transition is thus $U_c=0.80\cdot U_{\rm BR}=2.71D$, at which the charge gap closes and the QP coherence emerges with the D/H condensate simultaneously.
\begin{figure}
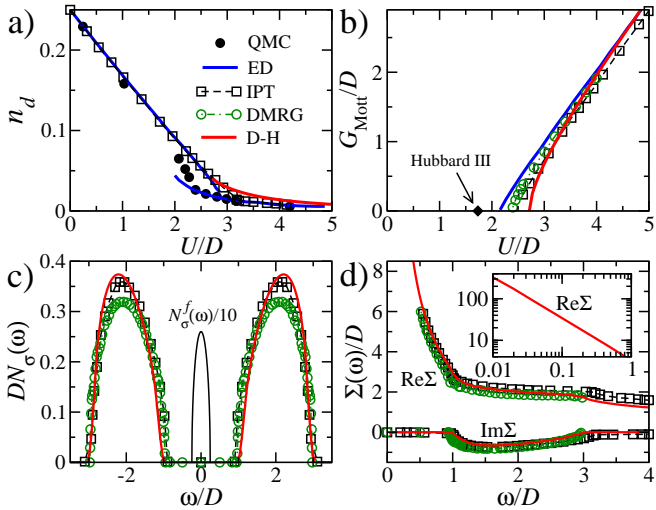

      \begin{center}
    \fig{3.4in}{fig3.eps}\caption{Comparison of the current theory (red lines) with the DMFT results (data from Ref.\cite{dmftrmp96,karski}) obtained using quantum monte carlo (QMC - solid black circles), exact diagonalization (ED - blue lines), iterative perturbation theory (IPT- open squares), and dynamical density matrix renormalization group (DMRG - open circles) as impurity solvers. (a) The doublon density as a function of $U$.
    (b) The Mott gap in the charge sector as a function of $U$. (c) The spectral density of states at $U=4D$. Thin solid line: spinon density of states. (d) The real and imaginary parts of the electron self-energy at $U=4D$. Inset: Real part of self energy on log-log plot, showing the $1/\omega$ dependence. \label{fig2}}
    \end{center}
    \vskip-0.5cm
    \end{figure}

In Figs ~3(a) and 3(b), the calculated doublon density and Mott gap are plotted in red solid lines as a function of $U/D$. Various single-site DMFT results \cite{dmftrmp96,karski,dv2012} are also plotted in Fig.~3 for comparison solely for the purpose to benchmark the results in the charge sector, despite of the different large-$z$ limit and the continuous Mott transition to a spin liquid at a single $U_c$.
The critical behavior of the Mott gap near $U_c$ can be obtained analytically from Eq.~(\ref{mottgap}), $G_{\rm Mott}(U)=\alpha\sqrt{U-U_c}$, $\alpha=2\sqrt{2t_b}$, where the square-root singularity is clearly seen in Fig.~2(b). Figs ~3(c) and 3(d) show the spectroscopic properties on the Mott insulating side benchmarked with corresponding DMFT results.
The local electron Green's function is obtained by convoluting those of the spinon and D/H ($Z$-boson) $G_\sigma(i\omega_n)=\sum_{i\nu_n}G_\sigma^f(i\omega_n-i\nu_n)G_Z(i\nu_n)$. The latter can be obtained readily from the spinon and the D/H Green's functions \cite{zhouwangwang}: $G_\sigma^f(i\omega_n)=\int d\epsilon \rho_0(\epsilon)G_\sigma^f(\epsilon,i\omega_n)$ and $G_Z(i\nu_n)=\int d\epsilon \rho_0(\epsilon)G_Z(\epsilon,i\nu_n)$. The electron spectral density is given by
$
N_\sigma(\omega)=-{1\over\pi} {\rm Im}G_\sigma(i\omega_n\to\omega+i0^+).
$
Fig.~3(c) shows $N_\sigma(\omega)$ obtained at $U=4D$, exhibiting the upper and the lower Hubbard bands separated by the Mott gap, in quantitative agreement with the DMFT results \cite{dmftrmp96,karski}. The spectral density of the spinons $N_\sigma^f(\omega)$ remains gapless as shown in Fig.~3(c) and contributes to thermodynamic properties of the spin liquid at low temperatures. We note in passing that these properties of the Mott transition/insulator are inaccessible to Gaussian fluctuations around the Kotliar-Ruckenstein saddle point for the putative BR transition at large $U$ \cite{raimondi,castellani}.
The central quantity in the large-$z$ limit is the local self-energy $\Sigma(\omega)$, which can be extracted by casting the local {\em electron} Green's function in the form
\begin{equation}
G_\sigma (\omega)=\int_{-D}^D d\epsilon \rho_0(\epsilon){1\over\omega-\epsilon-\Sigma(\omega)}.
\label{gdmft}
\end{equation}
Fig.~3(d) shows that the obtained $\Sigma(\omega)$ in the D-H binding theory is remarkably close to the real and imaginary part of the self-energy in DMFT at the same value of $U=4D$ \cite{dmftrmp96,karski}, including the scaling behavior ${\rm Re}\Sigma(\omega)\propto1/\omega$ inside the Mott gap shown in the inset.

The emergence of the spin-liquid Mott insulator with gapless spinon excitations requires the separation of spin and charge and is stable only if the gauge field that couples them is deconfining. To derive the gauge field action, we integrate out the matter fields by the hopping expansion \cite{im1995}. To leading order in $1/z$, the low energy effective gauge field action is obtained,
\begin{eqnarray}\label{action}
S_{\mathrm{eff}}
&=&-\frac{\eta}{z\pi^2}\sum_{\langle i,j\rangle} \int^{\beta}_0\!\!\mathrm{d}\tau_1\int^{\beta}_0\!\!\mathrm{d}\tau_2
\frac{\cos{(a_{ij}(\tau_1)-a_{ij}(\tau_2))}}{(\tau_1-\tau_2)^2}\nonumber\\
&&+\frac{1}{zC}\sum_{\langle i,j\rangle}\int^{\beta}_0\!\!\mathrm{d}\tau(\partial_{\tau}a_{ij})^2,
\end{eqnarray}
where the second term comes from integrating out the gapped D/H and corresponds to charging with the ``charging energy'' on a link $C\propto U^3/t_b^2$ in the large-$U$ limit. The first term with $\eta=1$, which is nonlocal in imaginary time and corresponds to dissipation, comes from the contribution from the gapless fermion spionons. It
is periodic in the gauge field consistent with its compact nature. Thus the gauge field action is dissipative. It has been argued under various settings that a large enough dissipation $\eta$ can drive the compact U(1) gauge field to the deconfinement phase at zero temperature \cite{nagaosa1993,wang2004,ksk2005}. In the large-$z$ limit, Eq.~(\ref{action}) shows that spatial correlations of the link gauge field are suppressed and the dissipative gauge field theory becomes local, i.e. $a_{ij}(\tau)=a(\tau)$. As a result, the action becomes identical to the dissipative tunneling action derived by Ambegaokar, Eckern, and Sch\"on \cite{aes1982} for a quantum dot coupled to metallic leads, or a Josephson junction with QP tunneling \cite{kampf}. The $2\pi$-periodicity of the compact gauge field requires $a(\tau)=\tilde a(\tau)+2\pi n \tau/\beta$ where $\tilde a(\tau)$ is single-valued and satisfies $\tilde a(0)=\tilde a(\beta)$, and $n$ is an integer winding number associated with charge quantization, i.e. the instantons in the electric field when charges tunnel in and out of the link. For a 2D array of dissipative tunnel junctions, it has been shown that there exists a confinement-deconfinement (C-DC) transition of the winding number at a critical $\eta_c^{2D}\simeq0.29$ \cite{mooji}. Using the Villain transformation \cite{villain}, one can show that the instanton action is described by a dissipative sine-Gordon model, exhibiting a C-DC transition at a critical dissipation $\eta_c=1/4$ \cite{longliang}. In our case, $\eta >\eta_c$, and the temporal proliferation of the instantons is suppressed by dissipation. Thus, the gauge electric field is deconfining and
the gapless U(1) spin-liquid is indeed the stable Mott insulating state.

In summary, we have provided an asymptotic solution of the Hubbard model in a novel large-$z$ limit and obtained a continuous Mott transition from a PM metal to a spin liquid Mott insulator where the opening of the Mott gap and the vanishing of the QP coherence coincide at the same critical $U_c$. We elucidated the essential role played by the D-H binding in such remarkable phenomena of strong correlation.
The present theory provides a concrete example for a gapless spin liquid Mott insulator where the spin-charge separation is realized in the deconfinement phase of the dissipative compact gauge field. The simplicity of the D-H binding theory for the Mott phenomena holds promise to become a calculational tool for studying Mott-Hubbard systems and materials with strong correlation.

We thank Y.P. Wang for useful discussions. This work is supported by the U.S. Department of Energy, Basic Energy Sciences Grant No. DE-FG02-99ER45747 (Z.W.), the Academy of Finland through its Centres of Excellence Programme (2015-2017) under project number 284621 (L.L.), and the Key Research Program of Frontier Sicences, CAS No. QYZDB-SSW-SYS012 (S.Z.). Z.W. thanks the hospitality of Aspen Center for Physics where this work was conceived, and the support of ACP NSF grant PHY-1066293.

\end{document}